\documentclass{article}

\usepackage{tablefootnote}

\usepackage{arxiv}
\usepackage[utf8]{inputenc}         
\usepackage[T1]{fontenc}            
\usepackage[hyphens]{xurl}          
\usepackage[hyphenbreaks]{breakurl}
\usepackage[hidelinks]{hyperref}    
\usepackage{booktabs}               
\usepackage{amsfonts}               
\usepackage{nicefrac}               
\usepackage{microtype}              
\usepackage{lipsum}
\usepackage{graphicx}
\graphicspath{ {images/} }
\usepackage{enumitem}
\usepackage{lscape}

\usepackage{subcaption}
\usepackage{float}
\usepackage{mathtools}

\usepackage{tikz}
\usepackage{tikz-dependency}
\usetikzlibrary{fit,positioning,arrows.meta,shapes}
\tikzset{neuron/.style={shape=circle, minimum size=1.25cm, 
  inner sep=0, draw, font=\small}, io/.style={neuron, fill=gray!20}}
\usetikzlibrary{shapes,arrows}
\usetikzlibrary{calc}
\usetikzlibrary{positioning}

\usepackage[symbol]{footmisc}

\usepackage{arydshln}
\usepackage{multirow}
\usepackage{soul}
\usepackage{adjustbox}
\usepackage{amsmath}
\usepackage{amsbsy}
\usepackage{amssymb}

\usepackage{algpseudocode}
\usepackage[ruled, linesnumbered]{algorithm2e}
\SetKwInput{Input}{Input}\SetKwInOut{Output}{Output}

\SetCommentSty{mycommfont}

\newlength{\commentWidth}
\newcommand{\atcp}[1]{\tcp*[f]{\makebox[\commentWidth]{#1\hfill}}}

\numberwithin{equation}{section}

\renewcommand{\thefootnote}{\arabic{footnote}}

\newcommand\blfootnote[1]{%
  \begingroup
  \renewcommand\thefootnote{}\footnote{#1}%
  \addtocounter{footnote}{-1}%
  \endgroup
}

\title{Using Intermarket Data to Evaluate the Efficient Market Hypothesis with Machine Learning}

\author{
    N'yoma Diamond \\
    Departments of Computer Science \& Data Science \\
    Worcester Polytechnic Institute \\
    Worcester, MA 01609 \\
    \texttt{badiamond@wpi.edu} \\
\And
    Grant Perkins \\
    Departments of Computer Science \& Data Science \\
    Worcester Polytechnic Institute \\
    Worcester, MA 01609 \\
    \texttt{gcperkins@wpi.edu} \\
}

\begin{document}

\maketitle

\begin{abstract}

In its semi-strong form, the Efficient Market Hypothesis (EMH) implies that technical analysis will not reveal any hidden statistical trends via intermarket data analysis. If technical analysis on intermarket data reveals trends which can be leveraged to significantly outperform the stock market, then the semi-strong EMH does not hold. In this work, we utilize a variety of machine learning techniques to empirically evaluate the EMH using stock market, foreign currency (Forex), international government bond, index future, and commodities future assets. We train five machine learning models on each dataset and analyze the average performance of these models for predicting the direction of future S\&P 500 movement as approximated by the SPDR S\&P 500 Trust ETF (SPY). From our analysis, the datasets containing bonds, index futures, and/or commodities futures data notably outperform baselines by substantial margins. Further, we find that the usage of intermarket data induce statistically significant positive impacts on the accuracy, macro F1 score, weighted F1 score, and area under receiver operating characteristic curve for a variety of models at the 95\% confidence level. This provides strong empirical evidence contradicting the semi-strong EMH.

\end{abstract}


\blfootnote{Code used in this paper is available at \url{https://github.com/Swiss-MQP-2022/Intermarket-ML-For-EMH}.}

\section{Introduction}

Forecasting the stock market is infamously difficult. Weak predictions are often explained by the Efficient Market Hypothesis (EMH). The EMH, in its semi-strong form, says that securities' values are determined by all publicly-available information~\cite{famaBehaviorStockMarketPrices1965, loAdaptiveMarketsFinancial2019, ruppertStatisticsFinanceIntroduction2004}. It is then implied by the semi-strong EMH that technical analysis cannot be used to outperform financial markets. In this paper, we define technical analysis as the process of evaluating statistical trends in market data. Should the semi-strong EMH hold, statistical trends should not reveal hidden information in market data, as all useful information for financial prediction should be immediately utilized to determine current price of a security~\cite{famaBehaviorStockMarketPrices1965}. However, what if looking at the foreign exchange, bond, and futures markets could reveal statistical trends in the stock market? If technical analysis on intermarket data could lead to significantly outperforming the stock market, then the semi-strong EMH would be challenged.

Numerous techniques exist for conducting technical analysis in finance. In contemporary research, machine learning is a particularly prevalent method. Using historical data, machine learning makes predictions about future data/outcomes. Access to vast quantities of information has never been simpler than in the Internet era. Economists, market makers, and individual investors can effortlessly access decades of returns data from markets such as the stock, foreign exchange (Forex), government bond, commodities future, and index future markets. These markets' data can be compiled into a set of intermarket datasets which can be used to train machine learning algorithms. Support vector machines (SVMs), logistic regression, decision trees, random forest, and k-nearest neighbors are notable examples of machine learning algorithms typically utilized for time series analysis. In this context, each model attempts to predict whether a stock price will rise or fall. The semi-strong EMH would be challenged if these models were used as a form of technical analysis on intermarket data to outperform models trained on stock market data.

In this work, We compile intermarket datasets from multiple markets for empirical analysis. We use these datasets to train machine learning models to predict the direction of movement in the stock market. In particular, we aim to predict the sign of movement (i.e., upward or downward) in the SPDR S\&P 500 Trust ETF (SPY)---an exchange-traded fund which closely track the S\&P 500 stock market index. In an effort to identify significant differences between multi-market technical analysis and stock market technical analysis, we compare the performance of several models across multiple metrics for each dataset. It is important to note that more data---i.e., data from multiple markets---does not necessarily result in improved performance, as excess data can lead to overfitting or spurious results. We challenge the semi-strong EMH in this paper by comparing the average model performances for each dataset, to see the impact that each market's data has on stock market prediction.

This paper is organized as follows: In Section~\ref{sec:background} we discuss existing literature pertaining to the EMH and financial prediction. In Section~\ref{sec:methodology} we discuss our experimental an analytical methodology. In Section~\ref{sec:results} we describe the results of our experiments and the analysis we conduct. In Section~\ref{sec:conclusion} we discuss our conclusions about the EMH based on our results and propose potential hypotheses for the behavior of our results. Lastly, in Section~\ref{sec:future work} we propose potential directions for future research based on the work in this paper.

\section{Background}\label{sec:background}

In order to contextualize our results and analyze their practical impact within the context of finance, we discuss the existing literature on financial forecasting, data analysis, and machine learning.

\subsection{The Efficient Market Hypothesis}

The Efficient Market Hypothesis (EMH)---developed in part by Eugene Fama in his 1965 dissertation and then popularized by a comprehensive review of existing research in 1970~\cite{famaBehaviorStockMarketPrices1965,famaEfficientCapitalMarkets1970}---states that markets are inherently efficient and rational. Fama categorizes market efficiency into three forms: Weak, semi-strong, and strong. Weak-form efficiency implies that the current market valuation of a given asset reflects all past pricing information for that asset. This means that technical analysis in a weak-form efficient market will be unsuccessful, and that fundamental analysis can only be used to predict the market in the short term. Semi-strong-form efficiency implies the the current market prices reflect all publicly available information. In a market that is semi-strong-form efficient, neither fundamental nor technical analysis (including intermarket analysis) can be used to predict market movement. Strong-form efficiency states that all information, including non-public information, is reflected in prices and that it is functionally impossible to predict the market~\cite{famaEfficientCapitalMarkets1970, loAdaptiveMarketsFinancial2019, ruppertStatisticsFinanceIntroduction2004}.

There is substantial disagreement among economists and academics regarding whether markets adhere to the EMH. While many reject some or all forms of the EMH, many also believe that attempts to predict market behavior are doomed for failure due to market efficiency. Over the past two decades, the publication of substantial literature identifying inconsistencies in the theory's expectations has eroded the general public's faith in the strong-form EMH, framing the theory as unrealistic. However, efforts to similarly disprove weak and semi-strong efficiency have yielded inconsistent results, leaving many scholars unsure about the EMH~\cite{yenEfficientMarketHypothesis2008, sewellEfficientMarketHypothesis2012, naseerEfficientMarketHypothesis2015}. 

\subsection{History of Financial Prediction}

The utilization of machine learning for predicting financial markets has been studied extensively. Machine learning encompasses techniques which utilize existing data to fit procedures for predicting outcomes based on a provided input~\cite{mitchellMachineLearning1990}. The literature offers various example applications of machine learning for stock market forecasting; many of which significantly outperform their more traditional counterparts. This was confirmed in more recent installments of the Makridakis Forecasting Competitions---which have been held roughly once a decade since 1982---and have the objective of comparing the accuracy of different forecasting methods. Prior to recent years, a recurring conclusion from these competitions has been that traditional, simpler methods are often able to perform similarly to their more complex counterparts. However, this changed at the latest editions of the competition series, M4 and M5, where a hybrid Exponential Smoothing Recurrent Neural Network method and LightGBM, won the competitions, respectively~\cite{wildiTimeSeriesApproach2022}.

To determine which machine learning methods might be effective for financial prediction, Milosevic~\cite{milosevicEquityForecastPredicting2018} trained various models, including support vector machine (SVM), random forest, logistic regression, naive Bayes, and Bayesian network. The models were trained to predict if the value of a particular stock would increase by 10\% in a year or not, given a variety of financial information about the company. The most successful model was random forest, with a score of 0.765 for precision, recall, and F-score~\cite{milosevicEquityForecastPredicting2018}. Huang et al.~\cite{huangForecastingStockMarket2005} also applied SVM to forecast the movement of the Nikkei 225 index. For their inputs they utilized the value of the S\&P 500 and the exchange rate between the US Dollar and Japanese Yen. They compared SVM to other classification models and concluded that SVM performed highly with an accuracy of 73\%, and attribute its success to its resistance to overfitting~\cite{huangForecastingStockMarket2005}. 

Wolff and Neugebauer~\cite{wolffTreeBasedMachineLearning2018} compared tree-based models with linear regression-type models for predicting equity markets. They found that more complex linear regression models, such as principal component regression and ridge regression, significantly outperformed tree-based models, with accuracies in the 67-68\% range compared to 58-67\% respectively. The models' returns on investments reflected these accuracies. The authors note that these results are likely due to a lower number of features and lower frequency of observations compared to other studies. This indicates the importance of both the quality and quantity of data used for training~\cite{wolffTreeBasedMachineLearning2018}. Leung et al.~\cite{leungMachineLearningApproach2014} used structural support vector machines to predict stock prices based on the financial data of collaborating companies. These models achieved a median testing accuracy of 58.644\%, leading them to conclude that the approach was effective~\cite{leungMachineLearningApproach2014}.

\section{Methodology}\label{sec:methodology}

To reach a meaningful conclusion, a broad dataset consisting of a number of assets is collected. We perform exploratory data analysis on this data to identify trends and any (un)desirable behavior. Experiments are conducted using a variety of machine learning models, each of which are trained using exhaustive hyperparameter searches and evaluated with a small suite of metrics. Baseline models and datasets are also included for comparative analysis. The effectiveness of these models is then computed and analyzed with respect to a number of metrics to identify the impact of varying intermarket asset types on predicting stock market returns.

\subsection{Gathering Intermarket Data}\label{sec:methodology-data}
Our full data set is composed of data from assets of following types: stocks, Forex, bonds, stock futures, and commodities futures. For our selected assets we chose to use daily pricing data with a desired start-date of January 1, 2000 and end-date of January 1, 2020. The features of each time series include date, opening price, closing price, high price, low price, volume (where available), and adjusted closing price (closing price adjusted for splits and dividends\footnote{More details available at \url{https://eodhistoricaldata.com/financial-apis/adjusted-close-and-close-whats-the-difference/}}). The exact assets we chose can be found in Table~\ref{table:finaldataset}~\cite{monetaryandeconomicdepartmentbankforinternationalsettlementsTriennialCentralBank2019, oecdOECDSovereignBorrowing2019}\footnote{Commodities futures were selected based on liquidity data from \url{www.cmegroup.com/markets/products.html}}. To predict on returns we performed a percent-change transformation on all features using the formula described in Equation~\ref{eq:percent-change},

\begin{equation}\label{eq:percent-change}
    x^r_t = \frac{x_t - x_{t-1}}{x_{t-1}}.
\end{equation}

In this equation, $x_t$ refers to the input data at time $t$, and $x^r_t$ represents the returns day-over-day returns at the same time-step (e.g., closing-to-closing returns, high-to-high returns, etc.). Note that, daily traded volume is also transformed using this equation, as a method of normalization.

\subsection{Exploratory Data Analysis}

Prior to dataset generation, we conducted exploratory data analysis to identify any notable characteristics of our selected assets and identify which assets to exclude. We searched for correlations between variables, assessed the normality of our data, and analyzed the presence of holes and outliers in our data. Normality was assessed visually using kernel density estimate plots~\cite{parzenEstimationProbabilityDensity1962, rosenblattRemarksNonparametricEstimates1956} and quantile-quantile plots~\cite{wilkProbabilityPlottingMethods1968}. Based on this analysis we believe our data is roughly close enough to normal that we techniques which operate better on normally distributed data could reasonably be used. We also utilized pairplots\footnote{As provided by the Seaborn Python library. See \url{https://seaborn.pydata.org/generated/seaborn.pairplot.html}.} and confusion matrices to understand correlation in our data. Most of our assets displayed weak positive correlation with SPY data. From this, we determined that all of our asset data is potentially valuable for predicting SPY returns. Lastly, we identified that holes and outliers were relatively minimal within our data set and could generally be ignored. However, many of our assets begin recording data later than our desired start date of January 1, 2000. Therefore, intermarket datasets utilizing these assets have some initial data ignored for alignment purposes (discussed in more detail in Section~\ref{sec:data gen}).

\subsection{Dataset Generation}\label{sec:data gen}
The assets selected for our experimentation datasets are displayed in Table~\ref{table:finaldataset}. We employ Algorithm~\ref{alg:dataset creation} to generate this collection of datasets. The first step of our algorithm is to create labels, and is shown in line 1 of Algorithm~\ref{alg:dataset creation}. We calculate the direction of the movement of S\&P 500 return, where +1 indicates upward movement and -1 indicates downward movement or no movement. Our labels are shifted forward one recorded day so that future returns are predicted. We then construct the power set ($\mathcal{P}$, line 3) of available asset types, excluding stocks. Each element of the power set then has S\&P 500 data inserted ($X$, line 4) so that all intermarket datasets contain historic S\&P 500 data. Because we are analyzing 4 intermarket asset types, this results in a total of 16 ($2^4$) unique asset type combinations. For each dataset generated we then align the prediction labels with the dataset by their available days of data. This removes all samples (days of data) from the dataset for which any of the component assets do not have data. Following alignment, standardization is applied, respresented by the \texttt{Standardize} function on line 5 of Algorithm~\ref{alg:dataset creation}. The formula used in this function is described in Equation~\ref{eq:standardization}. In this equation, $X$ refers to our input data, $\mu$ and $\sigma$ refer to the mean and standard deviation of the features of the dataset respectively, and $Z$ refers to the final standardized value,

\begin{equation}\label{eq:standardization}
    Z = \frac{X - \mu}{\sigma}.
\end{equation}

Finally, we prepare our datasets for use with our models by reshaping our data by input period. As represented by the \texttt{Window} function in line 5 of Algorithm~\ref{alg:dataset creation}, we create sliding windows on our data for use in prediction. In our experimentation, we use a window-size of 5 days. These periods are then split such that 80\% of each dataset was used for training (in-sample) while we used 20\% for testing (out-of-sample). 

\begin{table}[htbp]
\centering
\caption{Assets included in the final dataset}
\begin{tabular}{|l|l|l|} 
\hline
Asset Type                           & Asset                       & Data Start (Y-M-D)  \\ \hline\hline
Stock                                & SPY                         & 2000-01-03          \\ \hline
\multirow{6}{*}{Government Bond}     & US 5-Year                   & 2000-01-03          \\ 
                                     & US 10-Year                  & 2000-01-03          \\ 
                                     & US 30-Year\tablefootnote{United States 30-Year bond data are only available as futures from the EOD Historical Data API.}     & 2000-01-03          \\ 
                                     & Italy 10-Year               & 2010-01-04          \\ 
                                     & Japan 10-Year               & 2012-07-09          \\ 
                                     & United Kingdom 5-Year       & 2012-07-09          \\ \hline
\multirow{5}{*}{Forex}               & USD/CAD                     & 2000-01-03          \\ 
                                     & USD/EUR                     & 2000-01-03          \\ 
                                     & USD/GBP                     & 2000-01-04          \\ 
                                     & USD/JPY                     & 2000-01-03          \\ 
                                     & EUR/GBP                     & 2000-01-03          \\ \hline
\multirow{5}{*}{Index Futures}       & E-Mini SP 500 (ES)          & 2000-09-18          \\ 
                                     & EURO STOXX 50 (FESX)        & 2000-01-03          \\ 
                                     & Hang Seng Index (HSI)       & 2009-11-19          \\ 
                                     & Nikkei 225 (NK)             & 2007-10-12          \\ 
                                     & CBOE Volatility Index (VIX) & 2009-04-24          \\ \hline
\multirow{4}{*}{Commodities Futures} & Gold (GC)                   & 2000-08-30          \\ 
                                     & Natural Gas (NG)            & 2009-04-24          \\ 
                                     & Corn (ZC)                   & 2000-07-17          \\ 
                                     & Soy (ZS)                    & 2000-01-03          \\ \hline
\end{tabular}
\label{table:finaldataset}
\end{table}

\begin{algorithm}
\caption{Asset-based dataset creation}\label{alg:dataset creation}

\SetKwFunction{Align}{Align}
\SetKwFunction{Shift}{Shift}
\SetKwFunction{Standardize}{Standardize}
\SetKwFunction{Returns}{Returns}
\SetKwFunction{Sign}{Sign}
\SetKwFunction{Window}{Window}

\Input{S\&P 500 data $s$, list of intermarket asset types $I$}
\Output{List of asset-based datasets $D$}
\BlankLine
$Y \gets \Shift(\Sign(\Returns(s)))$ \atcp{Generate base market movement values $Y$} \\
$D \gets \varnothing$ \atcp{Initialize the list of datasets $D$} \\
\For(\atcp{For each possible subset of asset types (power set)}){$A \in \mathcal{P}(I)$} {
    $X \gets A \cup \{s\}$ \atcp{Set $X$ to subset of asset types and S\&P 500 data} \\
    $X \gets \Window(\Standardize(\Align(X, Y)))$ \atcp{Align, standardize, and window data} \\
    $D \gets D \cup \{X\}$ \atcp{Insert new dataset into $D$} \\
}
\Return{$D$}
\end{algorithm}

\subsection{Modeling}\label{sec:models}

We selected the following machine learning techniques for experimentation: Decision tree, random forest, logistic regression, support vector machine (SVM), and k-nearest neighbors. Each technique was chosen based on two criteria: Popularity (the technique is widely-used in the context of time series prediction and/or financial forecasting) and complexity (the technique is a simple and thoroughly-explored technique which is efficient to train and test). 

Each dataset in our collection is distinctly different from the next in the number of input features, samples, trends, and other qualities. As such, we believe that fixing model hyperparameters across all datasets would induce unfair (dis)advantages for models trained on different datasets. To mitigate this and ensure we are comparing the best possible models for each dataset, we perform exhaustive hyperparameter searches\footnote{As provided by the Scikit-Learn Python library. See \url{https://scikit-learn.org/stable/modules/generated/sklearn.model_selection.GridSearchCV.html}.} for each model-dataset pair. 

To limit any bias introduced by the hyperparameter searches, we employ 5-fold time-series-aware cross-validation\footnote{As provided by the Scikit-Learn Python library. See \url{https://scikit-learn.org/stable/modules/generated/sklearn.model_selection.TimeSeriesSplit.html}.} (CV) during hyperparameter searching. This ensures that the chosen hyperparameters for a model are truly optimal for the dataset based on extrapolation performance (as opposed to standard K-fold CV, which would optimize for interpolation). The hyperparameter search iterates through all possible combinations of the provided hyperparameters. For each available hyperparameter combination, macro F1 scores are computed for each CV fold and are averaged, creating a CV score. The search compares all of the observed CV scores and chooses the hyperparameter set which results in the highest score. Note that specific incompatible hyperparameter combinations are treated as having a CV score of zero.

\begin{table}[htbp]
\centering
\caption{Parameter grids used for hyperparameter grid-search}
\label{table:parameter grids}
\begin{tabular}{|l|l|} 
\hline
Classifier\tablefootnote{As provided by the Scikit-Learn Python library. See \url{https://scikit-learn.org/stable/modules/classes.html}}         & Parameter Grid                                                                                                                                                                                                \\ 
\hline \hline
DecisionTreeClassifier       & \begin{tabular}[c]{@{}l@{}}splitter=[best, random],\\ max\_depth=[5, 10, 25, None],\\ min\_samples\_split=[2, 5, 10, 50],\\ min\_samples\_leaf=[1, 5, 10]\end{tabular}                                    \\ 
\hline
RandomForestClassifier       & \begin{tabular}[c]{@{}l@{}}n\_estimators=[50, 100, 500],\\ criterion=[gini, entropy],\\ max\_depth=[5, 10, 25, None],\\ min\_samples\_split=[2, 5, 10, 50],\\ min\_samples\_leaf=[1, 5, 10]\end{tabular}  \\ 
\hline
LogisticRegression & \begin{tabular}[c]{@{}l@{}}penalty=[l1, l2],\\ C=[1e-3, 1e-2, 1e-1, 1, 1e+1, 1e+2, 1e+3],\\ solver=[newton-cg, lbfgs, liblinear]\end{tabular}                                                                         \\
\hline
LinearSVC          & \begin{tabular}[c]{@{}l@{}}penalty=[l1, l2],\\ C=[1, 4, 9, 16, 25],\\ loss=[hinge, squared\_hinge]\end{tabular}                                                                                       \\ 
\hline
KNeighborsClassifier                & \begin{tabular}[c]{@{}l@{}}n\_neighbors=[5, 10, 15, 20],\\ weights=[uniform, distance],\\ metric=[l1, l2, cosine]\end{tabular}                                                                      \\ 
\hline
\end{tabular}
\end{table}

For decision trees, we vary the splitter, max depth, minimum number of samples to split a node, and the minimum number of samples required to create a leaf. The splitter can be set to choose randomly or chose the most optimal split based on a metric (Gini impurity by default). The maximum depth controls how complex the tree can be, with None meaning to expand the tree with no depth limit. Two values control the splitting behavior of the tree: The minimum number of samples at which to split an internal node and the minimum number of samples required to create a leaf. Varying these values can significantly alter the shape of the tree and therefore its potential performance. 

For random forest we vary the hyperparameters identically to decision trees, with the addition of varying the number of estimators (the number of trees in the ensemble) to use. Varying the number of estimators allows us to adjust the importance of individual trees in the ensemble and potentially produce more uncommon results.

For logistic regression, we vary both the penalty method and regularization strength. Both parameters affect the model's resistance to overfitting. Certain penalty techniques and regularization strengths can also improve performance when applied correctly, so it is important to determine the correct regularization method for a given dataset. We also vary the solver to use, which impacts training behavior and how optimal performance is identified. 

For SVM, we vary the penalty method, regularization strength, and loss function. Varying the loss function changes the objective for which the SVM optimizes, thus changing its general behavior. Note that we exclusively utilize a linear kernel (i.e., decision boundary), as the running time of other kernels makes their use impractical.

For k-nearest neighbors, we varied the number of neighbors, the distance-weighting technique, and the distance metric. Increasing the number of neighbors reduces the chances of overfitting but decreases the precision of the model, making it difficult to identify highly localized behavior. Distance-weighting was also varied to either weigh all neighbors equally or weight them by distance (i.e., closer neighbors are more valuable). Lastly, distance calculation metrics were varied to consider any potential impact due to differing approaches to distance computation.

\subsection{Baselines Models for Comparison}\label{sec:baseline models}

To validate and contextualize the effectiveness of our models, we compared their performance with four baseline models: A random baseline, constant-prediction baseline, previous-value repetition, and a previous-value consensus. These baselines serve as simple, naive, and easily implementable reference points with which to compare our machine learning models. We assume that the presence of intermarket asset data should have no meaningful impact on the performance of these baseline models since they make no use of any provided intermarket asset data.

The random baseline produces a random prediction by sampling from a uniform distribution such that upward and downward movement are each predicted 50\% of the time. The random baseline is particularly notable in the financial investment context, as it has been identified previously that random strategies are often comparable to more complexly defined strategies~\cite{biondoAreRandomTrading2013}.

The constant-prediction baseline always predicts the class with the greatest number of overall occurrences in the training set. This is done by computing the training set's prior distribution. This baseline should inherently provide a minimum accuracy of 50\%, as any imbalance in returns direction will bias the baseline towards the more common outcome. 

The previous-value repetition baseline makes use of a naive repetition strategy. Specifically, the output of the baseline will be equivalent to the previous value observed. I.e., we predict upward movement if the previous day's S\&P 500 returns were positive, and downward movement if they were negative. In our results we abbreviate this baseline's name to ``Previous Baseline''

The previous-value consensus baseline operates by setting the predicted value to the consensus of the sign of returns from the previous $n$ days, where $n$ is some user-chosen number. I.e., this baseline will predict whichever movement was most common in the previous $n$ days. In our implementation we use $n=5$ days as this is also the window size we use for our datasets. This means that the baseline's prediction will be the consensus of the previous 5 days of data. Note that the previous-value repetition baseline is actually an instance of the consensus baseline where $n=1$. In our results we abbreviate this baseline's name to ``Consensus Baseline''

\subsection{Metrics for Analysis}\label{sec:metrics}

To analyze the performance of our models we use accuracy, macro-averaged F1 score, weighted F1 score, and area under receiver operator characteristic curve (AUC). These metrics provide simple numerical methods for evaluating and comparing classification model performance. While we prioritize macro F1 for hyperparameter selection, our analysis will broadly cover all of these metrics equally. Note that each of our metrics (aside from AUC) have a different bias due to class imbalance: Accuracy biases towards upwards movement (the most common class), macro F1 biases towards downwards movement, and weighted F1 is balanced and considered unbiased. This should provide us with meaningful insight into how our models and datasets perform, independent of class imbalances in our data.

\subsection{Experimentation and Analysis}\label{sec:experimentation and analysis}

Fifty replications are performed for each model-dataset pair. This provides us with a total sample size of 7650, or 50 replications times 9 models (four of which are our baseline models) times 17 datasets (one of which is the random data baseline). We record which model and dataset were used as well as the in- and out-of-sample metrics (as described in Section~\ref{sec:metrics}) observed for each model-dataset pair. The analysis of these metrics is split up into four parts: (1) Asset presence impact, (2) dataset performance, (3) model performance, and (4) analysis of variance (ANOVA) of asset type effects. 

Asset presence impact analysis focuses on the mean model performance in the presence or absence of each asset type. We empirically compare the mean observed performance depending on an individual asset type's presence to identify whether the inclusion of an asset improves or impairs model performance on average. To do this, for each asset type, we calculate the mean performance of all non-baseline models for datasets containing that asset type, and the mean for datasets not containing that asset type. We compare these results against the mean observed model performance on datasets containing only SPY data and datasets containing only random data. We only conduct this analysis on out-of-sample data.

Dataset performance analysis compares the mean performance of our non-baseline models with respect to each dataset. We calculate the mean with respect to each dataset of all non-baseline models' performances. We then use the mean performance metrics for each of our datasets to analyze whether specific combinations of asset types induce any notable improvement or degradation in model performance. This analysis is only conducted on out-of-sample data.

Model performance assesses the mean performance of each of our models across all of our datasets (excluding the random data baseline). This is done by computing the mean with respect to each model of the observed performance metrics on all non-random datasets. These results are compared to identify whether certain predictive models reliably outperform each other or our baseline models. This analysis is conducted separately on in-sample and out-of-sample data in order for us to identify and compare how our models behaved during training and provide potential explanations for observed behavior.

Lastly, factorial ANOVA models are generated with respect to asset type for each of our models. For each model, we have a $2\times4$ design such that each asset type is associated with a factor. An asset type's presence indicates the level of that factor, such that usage of Forex, bond, index futures, or commodities futures are our factors (4 factors) and an asset type being present in the dataset is the high-state for that asset type's factor, while its absence is its low-state (2 levels). We opt not to analyze interaction effects as we believe any impact due to the presence of a given asset type should be independent of any other asset type. Further, the presence of interactions makes it questionably meaningful to interpret the coefficients of the effects model, which we must do to identify whether the usage of a given intermarket asset type improves or impairs model performance. 

The intercepts of our ANOVA models can be interpreted to represent the mean performance in the absense of any intermarket assets (i.e., a model's mean performance on the dataset containing only SPY data) and the coefficients of each factor represents the average benefit of using that factor's asset type. The random data baseline and in-sample data were excluded from statistical analysis, as they do not provide meaningful information in this context. 

For each predictive model, two ANOVA models are generated, a full model containing all factor effects, and a reduced model containing only significant effects. This provides us with a suite of ANOVA models which we can use to identify whether the presence of a particular intermarket asset type induces a statistically significant change in model performance with respect to each of our metrics and models. Note that while full ANOVA models are computed, we do not analyze them in our results unless all effects in the full model were significant (and thus not requiring reduction). This is because the the presence of non-significant effects in an ANOVA model also impacts the other, potentially significant or non-significant effects. I.e., it is possible that removing one non-significant effect will result in another effect becoming significant or non-significant. The reduced models mitigate this fact by iteratively removing effects until only significant effects remain. The algorithm used to generate the full and reduced ANOVA models is described in Algorithm~\ref{alg:anova creation}.

\begin{algorithm}
\caption{ANOVA model generation}\label{alg:anova creation}

\SetKwData{Exclude}{exclude}
\SetKwData{Factors}{factors}
\SetKwData{Effects}{effects}

\SetKwFunction{ANOVA}{ANOVA}

\Input{Data to analyze $Y$, significance level $\alpha$}
\Output{Full ANOVA model $F$, Reduced ANOVA model $R$}
\BlankLine

$\Effects \gets Y.\Factors$ \atcp{Get \Effects for ANOVA (intermarket asset types)} \\
$R \gets \ANOVA(Y, \Effects)$ \atcp{Compute full ANOVA model on $Y$} \\ 
$F \gets R$ \atcp{Save the full model to $F$} \\
\While(\atcp{While an effect's $p$-value exceeds our significance level}){$\max(R_p) > \alpha$} {
    $\Effects \gets \Effects \setminus \{\arg\max(R_p)\}$ \atcp{Remove the least significant effect from \Effects} \\
    $R \gets \ANOVA(Y, \Effects)$ \atcp{Recompute ANOVA model $R$ using reduced \Effects} \\
}
\Return{$F, R$}
\end{algorithm}

This algorithm is applied once for each model-metric pair. I.e., we compute a full ANOVA model and a reduced ANOVA model for each response metric (described in Section~\ref{sec:metrics}) for each model (described in Sections~\ref{sec:models} and~\ref{sec:baseline models}). First, we store the $\textsf{effects}$ with respect to which we will analyze our metric-results $Y$. This is done by pulling the $\textsf{factors}$ of our experiment (i.e., our intermarket asset types), which is represented in line 1. A linear ANOVA model $R$ is then generated for $Y$ with respect to these effects (line 2). This first model is our full model, which is saved to $F$ and returned at the end of the algorithm (line 3). While the ANOVA model contains effects whose p-value ($R_p$) is greater than our desired significance level $\alpha$ (line 4), we find the the least significant effect ($\arg\max(R_p)$, line 5), remove it from our list of analyzed $\textsf{effects}$ (line 5), and recompute the ANOVA model $R$ (line 6). Once all remaining effects in the model have a p-value below our desired significance level, or no effects remain, the full model $F$ and the final reduced model $R$ are returned. In our experimentation, we use a significance level of $\alpha=0.05$ (i.e., a 95\% confidence level).

\section{Results}\label{sec:results}

Each of our models and datasets are evaluated using the methods described previously. Results of these experiments are used to analyze predictive performance with respect to intermarket asset presence, individual datasets (asset type combinations), and specific models. Statistical analysis is also conducted to identify the significance of each asset type on the performance of each model. In our results and analysis, asset types are codified by the first letter of the asset type's name. I.e., the presence of Forex, bond, index futures, and commodities futures in a dataset are represented by the letter codes F, B, I, and C respectively.

\subsection{Impact of Intermarket Asset Presence}\label{sec:individual asset impact}

In our experimentation, we observed that most of the analyzed intermarket asset types improved mean model performance when present. Numeric results can be seen in Table~\ref{tab:asset presence}. Note that asset types are not exclusive across rows of the table. This means that the set of datasets containing one specific asset type also includes datasets which may or may not have any of the \textit{other} asset types. For example, the row describing the presence of bond data represents the mean performance of \textit{all} datasets where bond data is present, including datasets containing other asset types as well. This also means that rows representing the absence of a particular asset contain the dataset with only S\&P 500 data (``SPY-Only''). No rows include the random data baseline except for the Random Data row. 

\begin{table}[htbp]
\caption{Mean model performance ($\pm\sigma$) with respect to the presence of each asset type}
\centering
\begin{tabular}{|lc|cccc|}
\hline
Asset Type & Presence & Accuracy & Macro F1 & Weighted F1 & ROC AUC \\
\hline\hline
\multirow[c]{2}{*}{Forex} & True & $0.531 \pm 0.029$ & $0.476 \pm 0.046$ & $0.496 \pm 0.036$ & $0.512 \pm 0.027$ \\
 & False & $0.539 \pm 0.027$ & $0.484 \pm 0.050$ & $0.504 \pm 0.040$ & $0.523 \pm 0.028$ \\ \hdashline
\multirow[c]{2}{*}{Bond} & True & $0.537 \pm 0.030$ & $\pmb{0.487 \pm 0.044}$ & $\pmb{0.506 \pm 0.035}$ & $0.521 \pm 0.031$ \\
 & False & $0.533 \pm 0.026$ & $0.473 \pm 0.051$ & $0.493 \pm 0.040$ & $0.514 \pm 0.025$ \\ \hdashline
\multirow[c]{2}{*}{Index Futures} & True & $0.538 \pm 0.029$ & $0.485 \pm 0.047$ & $0.505 \pm 0.038$ & $\pmb{0.524 \pm 0.032}$ \\
 & False & $0.532 \pm 0.027$ & $0.475 \pm 0.048$ & $0.495 \pm 0.037$ & $0.511 \pm 0.022$ \\ \hdashline
\multirow[c]{2}{*}{Commodities Futures} & True & $\pmb{0.539 \pm 0.031}$ & $0.484 \pm 0.048$ & $0.506 \pm 0.037$ & $0.520 \pm 0.033$ \\
 & False & $0.531 \pm 0.025$ & $0.476 \pm 0.048$ & $0.494 \pm 0.038$ & $0.515 \pm 0.023$ \\ \hline
SPY-Only & ~ & $0.529 \pm 0.024$ & $0.466 \pm 0.045$ & $0.483 \pm 0.034$ & $0.511 \pm 0.014$ \\
Random Data & ~ & $0.524 \pm 0.024$ & $0.462 \pm 0.055$ & $0.480 \pm 0.043$ & $0.502 \pm 0.019$ \\
\hline
\end{tabular}
\label{tab:asset presence}
\end{table}

The greatest mean improvement in accuracy was observed from commodities futures, resulting in an approximate 0.9\% increase on average. The greatest mean improvements to macro F1 and weighted F1 were observed from bonds, improving by roughly 0.013 for both. And the greatest mean improvement in AUC was observed from index futures, receiving an average 0.013 increase. Note that these asset-metric pairs, among all asset types analyzed, received the highest scores for their respective metrics when present, and the lowest scores for the same metrics when absent---excluding the S\&P 500 dataset (``SPY-Only'' and random data baseline datasets.

Models trained using intermarket asset data consistently outperformed compared to using only S\&P 500 data (``SPY-Only''). On average, models trained on SPY-Only had a mean accuracy of 52.9\% ($\sigma=3.1\%$), macro F1 of 0.466 ($\sigma=0.045$), weighted F1 of 0.480 ($\sigma=0.043$), and AUC of 0.511 ($\sigma=0.014$). Contrast this with the minimum values for each metric when considering the inclusion of other assets, which are 53.1\% (absence of commodities futures), 0.473 (absence of bonds), 0.493 (absence of bonds), and 0.511 (absence of index futures) for accuracy, macro F1, weighted F1, and AUC respectively. This suggests that, on average, the inclusion and analysis of intermarket assets improves model performance. This is also true when compared with the random data baseline dataset, which received the worst observed mean model performance. Random data received mean scores of 52.4\% ($\sigma=2.4\%$), 0.462 ($\sigma=0.055$), 0.480 ($\sigma=0.043$), and 0.502 ($\sigma=0.019$) for accuracy, macro F1, weighted F1, and AUC respectively. This puts performance on the random data baseline below all other datasets when averaged by asset presence presence, including the SPY-Only dataset, suggesting that even predicting on just S\&P 500 data alone is beneficial.

We observe that mean model performance consistently degrade across all metrics in the presence of Forex data, while all other asset types consistently improve. On average, the inclusion of Forex data results in decreases of 0.7\%, 0.008, 0.003, and 0.011 towards accuracy, macro F1, weighted F1, and AUC respectively. This suggests the inclusion of Forex data may be actively detrimental. This contrasts with bonds, index futures, and commodities futures, which all saw improvements in mean performance across all metrics due to their inclusion. However, the mean observed performance of datasets containing Forex data is still greater than the mean performance on SPY-Only. This is because the Forex presence row represents the average of all datasets containing Forex data, which also includes datasets containing other assets which may induce improvements in performance, outweighing potential degradation due to Forex presence.

\subsection{Impact of Individual Datasets}\label{sec:asset combinations}

In our experimentation, we observed that the vast majority of the analyzed datasets resulted in improvements to mean model performance when present, with a few notable exceptions. Numeric results can be seen in Table~\ref{tab:dataset results}. 

\begin{table}[htbp]
\caption{Mean performance of models ($\pm\sigma$) on each dataset (asset type combination)}
\centering
\begin{tabular}{|l|cccc|}
\hline
Asset Combination & Accuracy & Macro F1 & Weighted F1 & ROC AUC \\
\hline\hline
F & $0.520 \pm 0.021$ & $0.465 \pm 0.055$ & $0.481 \pm 0.043$ & $0.502 \pm 0.013$ \\
B & $0.537 \pm 0.022$ & $0.487 \pm 0.046$ & $0.504 \pm 0.035$ & $0.523 \pm 0.017$ \\
I & $0.534 \pm 0.022$ & $0.479 \pm 0.045$ & $0.496 \pm 0.036$ & $0.520 \pm 0.019$ \\
C & $0.532 \pm 0.028$ & $0.463 \pm 0.052$ & $0.489 \pm 0.038$ & $0.501 \pm 0.019$ \\
FB & $0.529 \pm 0.025$ & $0.480 \pm 0.042$ & $0.498 \pm 0.033$ & $0.514 \pm 0.024$ \\
FI & $0.540 \pm 0.025$ & $0.487 \pm 0.051$ & $0.503 \pm 0.043$ & $0.523 \pm 0.028$ \\
FC & $0.535 \pm 0.025$ & $0.461 \pm 0.052$ & $0.488 \pm 0.037$ & $0.508 \pm 0.020$ \\
BI & $0.533 \pm 0.023$ & $0.474 \pm 0.055$ & $0.493 \pm 0.044$ & $0.522 \pm 0.029$ \\
BC & $0.549 \pm 0.029$ & $0.505 \pm 0.039$ & $0.522 \pm 0.030$ & $0.528 \pm 0.025$ \\
IC & $0.542 \pm 0.025$ & $0.489 \pm 0.054$ & $0.509 \pm 0.041$ & $0.532 \pm 0.027$ \\
FBI & $0.524 \pm 0.028$ & $0.472 \pm 0.035$ & $0.490 \pm 0.027$ & $0.505 \pm 0.022$ \\
FBC & $0.524 \pm 0.032$ & $0.475 \pm 0.037$ & $0.494 \pm 0.028$ & $0.500 \pm 0.026$ \\
FIC & $0.532 \pm 0.033$ & $0.477 \pm 0.046$ & $0.497 \pm 0.034$ & $0.516 \pm 0.032$ \\
BIC & $\pmb{0.555 \pm 0.030}$ & $\pmb{0.509 \pm 0.044}$ & $\pmb{0.531 \pm 0.035}$ & $\pmb{0.547 \pm 0.041}$ \\
FBIC & $0.545 \pm 0.034$ & $0.494 \pm 0.033$ & $0.516 \pm 0.023$ & $0.526 \pm 0.035$ \\ \hdashline
SPY-Only & $0.529 \pm 0.024$ & $0.466 \pm 0.045$ & $0.483 \pm 0.034$ & $0.511 \pm 0.014$ \\
Random Data & $0.524 \pm 0.024$ & $0.462 \pm 0.055$ & $0.480 \pm 0.043$ & $0.502 \pm 0.019$ \\
\hline
\end{tabular}
\label{tab:dataset results}
\end{table}

The dataset containing bond, index futures, and commodities futures data (BIC) saw the best mean performance across all metrics. This dataset received a mean accuracy of 55.5\% ($\sigma=3.0\%$), macro F1 of 0.509 ($\sigma=0.044$), weighted F1 of 0.531 ($\sigma=0.035$), and AUC of 0.547 ($\sigma=0.041$). This aligns with our analysis in Section~\ref{sec:individual asset impact}, as these asset types all saw improvements in mean performance in their presence. 

The same four datasets held the top scores for all metrics. BIC produced the highest mean performance for all metrics. BC followed BIC for all metrics except AUC, for which IC held second place. FBIC was the third best for all metrics except AUC, for which BC held third place. Lastly, IC was the fourth best dataset for all metrics except ROC, where it was the second best.

Comparing against SPY-Only, we see that the vast majority of datasets outperform on average. In total, 11, 12, 14, and 10 datasets beat SPY-Only in terms of accuracy, macro F1, weighted F1, and AUC respectively (out of 15 datasets containing intermarket data). Of these datasets, 9 outperform compared to SPY-Only on all metrics: B, I, FI, BI, BC, IC, FIC, BIC, and FBIC. Notably, all but one of the datasets which underperformed contain Forex data. This coincides with our analysis in Section~\ref{sec:individual asset impact}, as we observed that datasets containing Forex data performed worse on average compared to datasets without.

\subsection{Model Performance}

For each model, performance metrics were averaged across all datasets. We use these results to identify how effectively the different models perform overall and compared to our baselines, ignoring the impact of individual assets or datasets. This also allows us to justify the meaningfulness of our models' predictions. 

\subsubsection{Out-of-Sample}\label{sec:out-of-sample}

To consider the predictive quality of our models, we summarize the performance of each model across non-baseline datasets (i.e., excluding the random sample dataset) when evaluated on out-of-sample data. These results can be seen in Table~\ref{tab:mean model performance test}.

\begin{table}[htbp]
\caption{Mean out-of-sample model performance ($\pm\sigma$), compared to baseline models}
\centering
\begin{tabular}{|l|cccc|}
\hline
Model & Accuracy & Macro F1 & Weighted F1 & ROC AUC \\
\hline\hline
Decision Tree & $0.511 \pm 0.024$ & $0.499 \pm 0.025$ & $0.508 \pm 0.023$ & $0.502 \pm 0.026$ \\
Random Forest & $0.536 \pm 0.020$ & $0.488 \pm 0.023$ & $0.509 \pm 0.021$ & $0.506 \pm 0.024$ \\
Logistic Regression & $0.537 \pm 0.021$ & $0.511 \pm 0.029$ & $0.525 \pm 0.026$ & $0.528 \pm 0.030$ \\
Linear SVM & $\pmb{0.570 \pm 0.012}$ & $0.401 \pm 0.030$ & $0.443 \pm 0.028$ & $\pmb{0.538 \pm 0.018}$ \\
KNN & $0.522 \pm 0.023$ & $0.501 \pm 0.024$ & $0.514 \pm 0.022$ & $0.513 \pm 0.024$ \\ \hdashline
Random Baseline & $0.516 \pm 0.026$ & $0.512 \pm 0.026$ & $0.518 \pm 0.026$ & $0.500 \pm 0.000$ \\
Constant Baseline & $0.566 \pm 0.008$ & $0.361 \pm 0.003$ & $0.409 \pm 0.010$ & $0.500 \pm 0.000$ \\
Previous Baseline & $0.536 \pm 0.021$ & $\pmb{0.527 \pm 0.021}$ & $\pmb{0.535 \pm 0.021}$ & - \\
Consensus Baseline & $0.496 \pm 0.010$ & $0.474 \pm 0.008$ & $0.504 \pm 0.011$ & - \\
\hline
\end{tabular}
\label{tab:mean model performance test}
\end{table}

In general, all of our models outperformed some of our baselines on one or more metrics. Specifically, decision trees beat 6 baseline-metric pairs, random forest beat 7, logistic regression beat 9, linear SVM beat 7, and k-nearest neighbors beat 7 (out of 14 total baseline-metric pairs). 

All of our models beat the baselines with respect to AUC. Both the random- and constant-prediction baselines received an AUC of 0.500, while AUC metrics could not meaningfully be calculated for the previous and consensus baselines due to our implementation. This is effectively the worst possible AUC. Decision trees and random forest only differ marginally from the baselines, receiving mean AUCs of 0.502 ($\sigma=0.026$) and 0.506 ($\sigma=0.024$) respectively. K-nearest neighbor and logistic regression models saw decent improvement in AUC, with mean scores of 0.513 ($\sigma=0.024$) and 0.528 ($\sigma=0.030$) respectively. Lastly, we observed substantial improvement from linear SVM models with a mean AUC of 0.538 ($\sigma=0.018$). Note that linear SVM was the only model whose mean minus one standard deviation is greater than 0.5. This is also true using two standard deviations.

Linear SVMs notably outperformed all other models and baselines in terms of accuracy and AUC, receiving scores of 57.0\% ($\sigma=1.2\%$) and 0.538 ($\sigma=0.018$) respectively. This beat the next-highest accuracy observed, which was 56.6\% ($\sigma=0.8\%$) from the constant-prediction baseline (i.e., always predicting upward movement). This suggests that linear SVMs may be quite effective at predicting the S\&P 500 using intermarket data. However, linear SVMs also had the worst mean macro and weighted F1 scores of our models. This continues to align with the constant-prediction baseline, which had the lowest mean macro and weighted F1 scores of the baseline models (though was still beaten by linear SVMs). 

Logistic regression beat the most baseline-metric pairs, having the most consistently high mean results and outperforming all predictive models on all metrics except for linear SVM's accuracy and AUC scores. While linear SVMs substantially beat all other models for accuracy and AUC, they also hadl the worst observed mean macro and weighted F1 scores of all predictive models, while logistic regression models were consistently effective across all metrics. Logistic regression had the highest observed mean macro and weighted F1 scores from non-baseline models, with 0.511 ($\sigma=0.029$) and 0.525 ($\sigma=0.026$) respectively. Logistic regression also had the next highest mean accuracy after linear SVMs.

It is worth noting that because these results average performance across all datasets, it is likely that poor performance on some datasets is counteracting good performance on other datasets. As described in Sections~\ref{sec:individual asset impact} and~\ref{sec:asset combinations}, we saw that datasets containing bonds, index futures, and/or commodities futures induced improvements on average, while the presence of Forex data is seemingly detrimental to model performance. This means that summarizing our results by model across all datasets may artificially penalize them in an otherwise mitigable manner. Even so, we observed that multiple models perform decently well and/or outperformed our baselines when averaged across all datasets.

\subsubsection{In-sample}

To consider the effectiveness of the training of our models, we summarize the performance of each model across non-baseline datasets (i.e., excluding the random sample dataset) when evaluated on in-sample data. These results can be seen in Table~\ref{tab:mean model performance train}.

\begin{table}[htbp]
\caption{Mean in-sample model performance ($\pm\sigma$), compared to baseline models}
\centering
\begin{tabular}{|l|cccc|}
\hline
Model & Accuracy & Macro F1 & Weighted F1 & ROC AUC \\
\hline\hline
Decision Tree & $0.831 \pm 0.099$ & $0.827 \pm 0.103$ & $0.829 \pm 0.101$ & $0.899 \pm 0.085$ \\
Random Forest & $\pmb{0.998 \pm 0.010}$ & $\pmb{0.998 \pm 0.010}$ & $\pmb{0.998 \pm 0.010}$ & $\pmb{1.000 \pm 0.002}$ \\
Logistic Regression & $0.639 \pm 0.058$ & $0.623 \pm 0.068$ & $0.630 \pm 0.066$ & $0.686 \pm 0.077$ \\
Linear SVM & $0.571 \pm 0.023$ & $0.409 \pm 0.046$ & $0.441 \pm 0.045$ & $0.675 \pm 0.144$ \\
KNN & $0.711 \pm 0.113$ & $0.705 \pm 0.115$ & $0.709 \pm 0.113$ & $0.756 \pm 0.098$ \\ \hdashline
Random Baseline & $0.501 \pm 0.008$ & $0.500 \pm 0.008$ & $0.503 \pm 0.008$ & $0.500 \pm 0.000$ \\
Constant Baseline & $0.551 \pm 0.007$ & $0.355 \pm 0.003$ & $0.392 \pm 0.009$ & $0.500 \pm 0.000$ \\
Previous Baseline & $0.478 \pm 0.009$ & $0.472 \pm 0.010$ & $0.478 \pm 0.009$ & - \\
Consensus Baseline & $0.507 \pm 0.007$ & $0.497 \pm 0.007$ & $0.510 \pm 0.007$ & - \\
\hline
\end{tabular}
\label{tab:mean model performance train}
\end{table}

In contrast to the out-of-sample performance described previously, the in-sample performance of every model significantly outperforms all baselines across all metrics with only two exceptions. Random Forest in particular saw the highest in-sample performance, predicting nearly perfectly on in-sample data. Random Forest received a mean accuracy of 99.8\% ($\sigma=1.0\%$), both macro and weighted F1 of both 0.998 ($\sigma=0.010$), and AUC of 1.000 ($\sigma=0.002$). Considering the predictive models' out-of-sample performance, the most likely explanation for this is overfitting due to the extreme complexity of random forest, being an ensemble technique utilizing decision trees, which are also prone to overfitting due to potentially excessive complexity. 

By contrast, the worst models in terms of in-sample performance were our best performing models with respect to out-of-sample performance. Specifically, logistic regression and linear SVM models received mean accuracy, macro F1, weighted F1, and AUC scores of 63.9\% ($\sigma=5.8\%$) and 57.1\% ($\sigma=2.3\%$); 0.623 ($\sigma=0.068$) and 0.409 ($\sigma=0.046$); and 0.630 ($\sigma=0.066$) and 0.441 ($\sigma=0.045$); and 0.686 ($\sigma=0.077$) and 0.675 ($\sigma=0.144$) respectively. This aligns with our assertion about overfitting, as SVM and logistic regression models are generally regarded as being highly resistant to overfitting~\cite{nandiConditionMonitoringVibration2019}. Linear SVMs are also the only model to not beat all of the baselines across all metrics, receiving lower mean macro and weighted F1 scores compared to all of the baselines except the constant baseline.

\subsection{Statistical Analysis}

Using ANOVA as described in Section~\ref{sec:experimentation and analysis}, we observed that all non-baseline models saw statistically significant impacts from the presence of one or more intermarket assets all metrics. The ANOVA models are analyzed on a per-model basis, allowing us to disentangle impact of individual asset types from the impact of differing models. Note that we denote the omission of an effect in the reduced ANOVA models by putting a dash (-) in the cell for that effect. E.g., if bonds (B) did not result in a statistically significant impact on accuracy, the coefficient and p-value associated with bonds for accuracy will display a dash, as the effect of bonds is omitted from that model due to not being significant.

Note that statistically significant effects with a negative coefficient can be practically ignored. This is because if an asset type induces a negative impact on a predictive model's performance, we can simply omit that asset type from the model. That said, we still include these asset types in the ANOVA models due to their significance and to gain knowledge about how the different asset types impact different models---positively or negatively.

The reduced ANOVA models for decision trees (visible in Table~\ref{tab:dec tree anova}) were the smallest of all of the predictive models we analyzed. At the 95\% confidence level, we observe that index futures have a statistically significant impact on decision tree accuracy ($p=0.002$), resulting in an average improvement of 0.264\% when present. Further, bonds have a statistically significant impact on weighted F1 ($p=0.021$), resulting in an average improvement of 0.00190 when present. This suggests that the inclusion of these asset types may provide meaningful improvements to the effectiveness of decision tree models. However, the only statistically significant effects for macro F1 and AUC came from commodities futures, which had a negative impact in both cases. As a result, we fail to identify whether the usage of intermarket asset data improves the macro F1 and AUC of decision tree models. That said, it is important to remind that, as seen in Section~\ref{sec:out-of-sample}, decision tree models were generally much less successful than our other models in the first place.

\begin{table}[htbp]
\centering
\caption{ANOVA models for decision tree performance}
\begin{tabular}{|l|rc|rc|rc|rc|}
\hline
 & \multicolumn{2}{|c|}{Accuracy} & \multicolumn{2}{|c|}{Macro F1} & \multicolumn{2}{|c|}{Weighted F1} & \multicolumn{2}{|c|}{ROC AUC} \\
Effect & Coefficient & P-value & Coefficient & P-value & Coefficient & P-value & Coefficient & P-value \\
\hline\hline
F & - & - & - & - & - & - & - & - \\
B & - & - & - & - & $0.00190$ & $0.021$ & - & - \\
I & $0.00264$ & $0.002$ & - & - & - & - & - & - \\
C & - & - & $-0.00206$ & $0.018$ & - & - & $-0.00237$ & $0.011$ \\ \hdashline
Intercept & $0.51067$ & - & $0.49941$ & - & $0.50829$ & - & $0.50217$ & - \\
\hline
\end{tabular}
\label{tab:dec tree anova}
\end{table}

For random forest models (Table~\ref{tab:random forest anova}), many more intermarket asset types display statistically significant impacts on predictive performance compared to decision trees. For accuracy, both bonds and commodities futures display statistically significant positive impacts. On average, bonds resulted in a 0.457\% increase ($p=0.000$) in prediction accuracy, and commodities futures induced a 0.225\% increase ($p=0.001$). Bonds also induced statistically significant improvement towards weighted F1 and AUC, with average benefits of 0.00228 ($p=0.002$) and 0.00229 ($p=0.004$) respectively. In line with decision trees, none of the statistically significant effects improved macro F1. As a result, while we have statistically significant evidence to justify improvements in accuracy, weighted F1, and AUC due to the usage of intermarket assets, we have insufficient evidence to justify improvements in macro F1. That said, recall that random forest models had the worst mean macro F1 observed aside from linear SVMs, which notably underperformed in their own right.

\begin{table}[htbp]
\centering
\caption{ANOVA models for random forest performance}
\begin{tabular}{|l|rc|rc|rc|rc|}
\hline
 & \multicolumn{2}{|c|}{Accuracy} & \multicolumn{2}{|c|}{Macro F1} & \multicolumn{2}{|c|}{Weighted F1} & \multicolumn{2}{|c|}{ROC AUC} \\
Effect & Coefficient & P-value & Coefficient & P-value & Coefficient & P-value & Coefficient & P-value \\
\hline\hline
F & - & - & $-0.00314$ & $0.000$ & $-0.00260$ & $0.000$ & $-0.00204$ & $0.014$ \\
B & $0.00457$ & $0.000$ & - & - & $0.00228$ & $0.002$ & $0.00229$ & $0.006$ \\
I & - & - & $-0.00286$ & $0.000$ & $-0.00193$ & $0.010$ & $-0.00224$ & $0.007$ \\
C & $0.00225$ & $0.001$ & $-0.00211$ & $0.009$ & - & - & $-0.00227$ & $0.007$ \\ \hdashline
Intercept & $0.53550$ & - & $0.48815$ & - & $0.50863$ & - & $0.50551$ & - \\
\hline
\end{tabular}
\label{tab:random forest anova}
\end{table}

With respect to logistic regression models, our experimentation showed that all asset types and their interactions were statistically significant. Further, all metrics have two or three statistically significant positive effects due to intermarket data. Looking at the ANOVA results in Table~\ref{tab:logistic regression anova} we see that index and commodities futures induce a statistically significant positive impact on model performance with respect to all metrics. Similarly, bonds improve model performance with on all metrics except accuracy. All of these effects received p-values of effectively 0, with the exception of bond data's impact on AUC, which received a p-value of 0.006. Notably, index futures data induced rather substantial improvements in all metrics, resulting in an average increase of 1.042\%, 0.01805, 0.01591, and 0.02094 for accuracy, macro F1, weighted F1, and AUC respectively. Logistic regression was one of our best-performing predictive models on average, as identified in Section~\ref{sec:out-of-sample}. It is likely that logistic regression's high mean performance can be explained by the positive impacts observed from bond, index futures, and commodities futures data. The only effect which induces a significant negative impact is Forex data (with the exception of accuracy, for which bonds also induce a negative impact). This aligns with our observations in Section~\ref{sec:individual asset impact}.

\begin{table}[htbp]
\centering
\caption{ANOVA models for logistic regression performance}
\begin{tabular}{|l|rc|rc|rc|rc|}
\hline
 & \multicolumn{2}{|c|}{Accuracy} & \multicolumn{2}{|c|}{Macro F1} & \multicolumn{2}{|c|}{Weighted F1} & \multicolumn{2}{|c|}{ROC AUC} \\
Effect & Coefficient & P-value & Coefficient & P-value & Coefficient & P-value & Coefficient & P-value \\
\hline\hline
F & $-0.00661$ & $0.000$ & $-0.00139$ & $0.046$ & $-0.00236$ & $0.000$ & $-0.00947$ & $0.000$ \\
B & $-0.00263$ & $0.000$ & $0.00811$ & $0.000$ & $0.00544$ & $0.000$ & $0.00183$ & $0.006$ \\
I & $0.01042$ & $0.000$ & $0.01805$ & $0.000$ & $0.01591$ & $0.000$ & $0.02094$ & $0.000$ \\
C & $0.00335$ & $0.000$ & $0.00928$ & $0.000$ & $0.00926$ & $0.000$ & $0.00382$ & $0.000$ \\ \hdashline
Intercept & $0.53654$ & - & $0.51143$ & - & $0.52512$ & - & $0.52818$ & - \\
\hline
\end{tabular}
\label{tab:logistic regression anova}
\end{table}

Similarly to logistic regression, linear SVMs (Table~\ref{tab:svm anova}) saw statistically significant positive impacts from the presence of bond, index futures, and commodities futures data across all metrics (all with $p=0.000$). Also similarly, the only effect which induces a significant negative impact on performance is Forex data. However, unlike with logistic regression, this negative impact is only significant for accuracy and AUC, and bonds induce a statistically significant positive impact on accuracy. This aligns with and provides potential explanation for our observations in Sections~\ref{sec:asset combinations} and~\ref{sec:out-of-sample}. Given the positive impact of most of the assets, it is likely that we can attribute the success of linear SVM models to these assets. Interestingly, the metrics for which Forex data does not have a significant negative impact---macro and weighted F1---are also the metrics which linear SVM notably underperformed in. 

\begin{table}[htbp]
\centering
\caption{ANOVA models for linear SVM performance}
\begin{tabular}{|l|rc|rc|rc|rc|}
\hline
 & \multicolumn{2}{|c|}{Accuracy} & \multicolumn{2}{|c|}{Macro F1} & \multicolumn{2}{|c|}{Weighted F1} & \multicolumn{2}{|c|}{ROC AUC} \\
Effect & Coefficient & P-value & Coefficient & P-value & Coefficient & P-value & Coefficient & P-value \\
\hline\hline
F & $-0.00090$ & $0.000$ & - & - & - & - & $-0.00222$ & $0.000$ \\
B & $0.00335$ & $0.000$ & $0.01900$ & $0.000$ & $0.01730$ & $0.000$ & $0.00685$ & $0.000$ \\
I & $0.00227$ & $0.000$ & $0.00858$ & $0.000$ & $0.00762$ & $0.000$ & $0.01325$ & $0.000$ \\
C & $0.00886$ & $0.000$ & $0.00952$ & $0.000$ & $0.01375$ & $0.000$ & $0.00447$ & $0.000$ \\ \hdashline
Intercept & $0.56966$ & - & $0.40082$ & - & $0.44272$ & - & $0.53837$ & - \\
\hline
\end{tabular}
\label{tab:svm anova}
\end{table}

K-nearest neighbors (Table~\ref{tab:knn anova}) displayed the same significant effects across all metrics, with the exception of index futures. Specifically, bond and commodities futures data all induced statistically significant positive impacts on performance for all metrics, while Forex data induced significant negative impacts, all with p-values of effectively 0. Index futures were not significant for any metric with respect to k-nearest neighbors models.

\begin{table}[htbp]
\centering
\caption{ANOVA models for k-nearest neighbors performance}
\begin{tabular}{|l|rc|rc|rc|rc|}
\hline
 & \multicolumn{2}{|c|}{Accuracy} & \multicolumn{2}{|c|}{Macro F1} & \multicolumn{2}{|c|}{Weighted F1} & \multicolumn{2}{|c|}{ROC AUC} \\
Effect & Coefficient & P-value & Coefficient & P-value & Coefficient & P-value & Coefficient & P-value \\
\hline\hline
F & $-0.01109$ & $0.000$ & $-0.01420$ & $0.000$ & $-0.01315$ & $0.000$ & $-0.01288$ & $0.000$ \\
B & $0.00278$ & $0.000$ & $0.00479$ & $0.000$ & $0.00482$ & $0.000$ & $0.00470$ & $0.000$ \\
I & - & - & - & - & - & - & - & - \\
C & $0.00867$ & $0.000$ & $0.00484$ & $0.000$ & $0.00707$ & $0.000$ & $0.00787$ & $0.000$ \\ \hdashline
Intercept & $0.52211$ & - & $0.50128$ & - & $0.51397$ & - & $0.51297$ & - \\
\hline
\end{tabular}
\label{tab:knn anova}
\end{table}

Looking at the random baseline (Table~\ref{tab:random baseline anova}), we see that bonds and commodities futures induce statistically significant impacts on accuracy, macro F1, and weighted F1, all with p-values of effectively 0. This contradicts our assumption in Section~\ref{sec:baseline models} that the presence of intermarket data should have no impact on baseline model performance. That said, we failed to observe any statistically significant effects on AUC due to intermarket asset presence, aligning with our assumptions.

\begin{table}[htbp]
\centering
\caption{ANOVA models for random baseline performance}
\begin{tabular}{|l|rc|rc|rc|rc|}
\hline
 & \multicolumn{2}{|c|}{Accuracy} & \multicolumn{2}{|c|}{Macro F1} & \multicolumn{2}{|c|}{Weighted F1} & \multicolumn{2}{|c|}{ROC AUC} \\
Effect & Coefficient & P-value & Coefficient & P-value & Coefficient & P-value & Coefficient & P-value \\
\hline\hline
F & - & - & - & - & - & - & - & - \\
B & $0.00869$ & $0.000$ & $0.00842$ & $0.000$ & $0.00873$ & $0.000$ & - & - \\
I & - & - & - & - & - & - & - & - \\
C & $0.00897$ & $0.000$ & $0.00814$ & $0.000$ & $0.00936$ & $0.000$ & - & - \\ \hdashline
Intercept & $0.51636$ & - & $0.51229$ & - & $0.51819$ & - & $0.50000$ & - \\
\hline
\end{tabular}
\label{tab:random baseline anova}
\end{table}

Looking at the constant baseline (Table~\ref{tab:constant baseline anova}), we see that bonds, index futures, and commodities futures induce statistically significant impacts on accuracy, macro F1, and weighted F1, and that Forex data induce a statistically significant impact on AUC. This is similar to what we observed earlier with the random baseline, also all having p-values of effectively 0, with the exception of Forex data's impact on AUC, which had a p-value of 0.026.

\begin{table}[htbp]
\centering
\caption{ANOVA models for constant baseline performance}
\begin{tabular}{|l|rc|rc|rc|rc|}
\hline
 & \multicolumn{2}{|c|}{Accuracy} & \multicolumn{2}{|c|}{Macro F1} & \multicolumn{2}{|c|}{Weighted F1} & \multicolumn{2}{|c|}{ROC AUC} \\
Effect & Coefficient & P-value & Coefficient & P-value & Coefficient & P-value & Coefficient & P-value \\
\hline\hline
F & - & - & - & - & - & - & $0.00000$ & $0.026$ \\
B & $0.00180$ & $0.000$ & $0.00074$ & $0.000$ & $0.00212$ & $0.000$ & - & - \\
I & $0.00041$ & $0.000$ & $0.00016$ & $0.000$ & $0.00050$ & $0.000$ & - & - \\
C & $0.00736$ & $0.000$ & $0.00300$ & $0.000$ & $0.00872$ & $0.000$ & - & - \\ \hdashline
Intercept & $0.56610$ & - & $0.36146$ & - & $0.40930$ & - & $0.50000$ & - \\
\hline
\end{tabular}
\label{tab:constant baseline anova}
\end{table}

The previous baseline (Table~\ref{tab:previous baseline anova}) displays effectively identical behavior to the constant baseline in terms of effect significance. we see that bonds, index futures, and commodities futures induce statistically significant impacts on accuracy, macro F1, and weighted F1. Since AUC is not relevant to the previous baseline, Forex displays no statistically significant impact on any metrics with respect to the previous baseline.

\begin{table}[htbp]
\centering
\caption{ANOVA models for previous baseline performance}
\begin{tabular}{|l|rc|rc|rc|}
\hline
 & \multicolumn{2}{|c|}{Accuracy} & \multicolumn{2}{|c|}{Macro F1} & \multicolumn{2}{|c|}{Weighted F1} \\
Effect & Coefficient & P-value & Coefficient & P-value & Coefficient & P-value \\
\hline\hline
F & - & - & - & - & - & - \\
B & $0.01519$ & $0.000$ & $0.01500$ & $0.000$ & $0.01516$ & $0.000$ \\
I & $0.00929$ & $0.000$ & $0.00931$ & $0.000$ & $0.00927$ & $0.000$ \\
C & $0.00414$ & $0.000$ & $0.00229$ & $0.000$ & $0.00410$ & $0.000$ \\ \hdashline
Intercept & $0.53557$ & - & $0.52733$ & - & $0.53549$ & - \\
\hline
\end{tabular}
\label{tab:previous baseline anova}
\end{table}

The consensus baseline (Table~\ref{tab:consensus baseline anova}) is the only baseline to display statistically significant negative effects. While the consensus baseline is also seemingly unaffected by Forex data, bonds and index futures data appear to be detrimental towards the consensus baseline, while commodities futures data improve it.

\begin{table}[htbp]
\centering
\caption{ANOVA models for consensus baseline performance}
\begin{tabular}{|l|rc|rc|rc|}
\hline
 & \multicolumn{2}{|c|}{Accuracy} & \multicolumn{2}{|c|}{Macro F1} & \multicolumn{2}{|c|}{Weighted F1} \\
Effect & Coefficient & P-value & Coefficient & P-value & Coefficient & P-value \\
\hline\hline
F & - & - & - & - & - & - \\
B & $-0.00157$ & $0.000$ & $-0.00320$ & $0.000$ & $-0.00112$ & $0.000$ \\
I & $-0.00145$ & $0.000$ & $-0.00110$ & $0.000$ & $-0.00185$ & $0.000$ \\
C & $0.00855$ & $0.000$ & $0.00537$ & $0.000$ & $0.00889$ & $0.000$ \\ \hdashline
Intercept & $0.49597$ & - & $0.47412$ & - & $0.50353$ & - \\
\hline
\end{tabular}
\label{tab:consensus baseline anova}
\end{table}

\section{Conclusion}\label{sec:conclusion}

In our experimentation, we found substantial evidence to suggest that the usage of intermarket data improves the prediction performance of machine learning models. We observed that, on average, the usage of bonds, index futures, and/or commodities futures assists machine learning models in predicting the future direction of S\&P 500 movement. Further, we identified a large number of varying combinations of intermarket assets which induced substantial improvements in the mean prediction performance of a variety of machine learning models. Most notably, we found that bonds, index futures, and commodities futures data induced statistically significant improvements in the performance of the majority of our machine learning models across nearly all metrics. We believe that this outcome is due to the inherent association bonds and futures have with a specified time to maturity. I.e., their prices fluctuates with respect to an expected \textit{future} value or yield. This fact in conjunction with their general correlation with the S\&P 500 provides probable evidence to explain why these assets improve model performance.

On average, we observed the inclusion of commodities futures data provides a very notable positive impact on the predictive capabilities of our models. Additionally, a number of specific combinations of intermarket assets consistently saw notably high performance over baselines. Specifically, datasets BIC, BC, FBIC, and IC consistently outperformed on all metrics, suggesting that these datasets may prove particularly relevant towards the task of improving prediction performance on the S\&P 500. Note that all of these datasets contain commodities futures data. This provides further evidence suggesting that the inclusion of commodities futures data provides a substantially beneficial impact on prediction quality.

While bonds, commodities futures, and index futures often improved model performance, we noticed across a number of instances that the usage of Forex data was detrimental towards model performance. On average, the presence of Forex data consistently made models worse, and the usage of Forex data had a statistically significant negative impact on nearly all metrics for all predictive models. A possible explanation for this is that Forex data may not be sufficiently correlated with the S\&P 500 to extract any useful information for prediction. As a result, models may likely be overfitting on the Forex data, resulting in performance degradation. We believe this difference occurs due to bonds and futures sharing the characteristic of having a defined time to maturity. Specifically, Forex data lacks the specified time to maturity that characterizes bonds and futures, instead simply corresponding to the exchange rate at the exact moment of sampling. 

Additionally, due to the statistical significance of the various intermarket assets on our baseline model performance, we believe our experimental design may unintentionally introduce a confounding effect on performance due to how data alignment is conducted. Specifically, in order to maximize the amount of data available for training, each dataset was aligned separately to contain the maximum available number of samples. As a result, the different datasets contained data for different days, which may have substantially impacted results for each dataset. That said, we believe this fact is mostly mitigated by the use of averaging across datasets to identify the mean impact of individual assets independent of dataset. However, this may be a valuable point of consideration in future research.

All considered, we have found substantial statistically significant evidence to suggest that the usage of intermarket data can be used to improve the ability of machine learning models to predict future movement of the S\&P 500. This contradicts the semi-strong form of the efficient market hypothesis, which suggests that such analysis should not be beneficial. Specifically, we show empirically that the inclusion of bonds, index futures, and commodities futures data is greatly beneficial.

\section{Future Work}\label{sec:future work}

There are multiple aspects of this work that could be expanded upon. This being the case, we would like to propose a number of potential directions for future research based on our results.

As mentioned in Section~\ref{sec:conclusion}, aligning each dataset individually may introduce a confounding effect when trying to identify the impact of intermarket data. In order to mitigate this, future work should aim to align all datasets to only contain data for the same days.

As stated in Section~\ref{sec:methodology-data}, our data only consisted of a handful of features for each asset. As a result we may have omitted features which could have proven beneficial towards our models. As an example, in our experimentation we excluded yield data from our bonds, which may have had a substantial impact had it been included~\cite{angWhatDoesYield2006,wrightYieldCurvePredicting2006}. Further, the availability of our data was heavily limited by the API from which we sourced our data. This being the case, it may be beneficial to consider using alternative data sources which may have higher quality data with more features/metrics. 

Further work may benefit from investigating the use of more complex approaches. Specifically, deep learning models---such as a neural networks or long short-term memory---or ensemble approaches--such as bucket of models techniques---may benefit differently from intermarket data. 

Another development would be to apply regression instead of classification to predict exact future returns. Regression may be impacted differently by intermarket data compared to classification. This could also allow for the evaluation of predictive risk assessment.

As stated in Section~\ref{sec:data gen}, we utilized a window size of 5 days. However, larger or smaller window sizes may result in different impacts on model performance due to containing different amounts of potentially useful information. Additionally, non-linear windows may be considered such that the relative amount of samples in the window is proportional to its age (i.e., older data is provided in smaller quantities).

In this paper we did not explore the usage of transformations on our input data. However, transforming model inputs may be beneficial towards prediction quality. Specifically, linear discriminant analysis may prove beneficial towards the task of classification and Fourier analysis may help to identify underlying semi-cyclic behavior, which may be useful towards market forecasting.

In our work we utilized data polled on a daily basis. However, the impact of intermarket data on predicting market movement may vary depending on the polling basis. It may be the case that the value of intermarket data on forecasting capabilities is different between intraday, daily, weekly, monthly, or quarterly bases. This is especially valuable as data pertaining to certain assets---notably bond yields~\cite{angWhatDoesYield2006,wrightYieldCurvePredicting2006}---are frequently used as indicators of future long-term economic changes. 

Considerations could also be given to the relative value of intermarket asset data on predicting varying distances into the future. It is possible that intermarket data is more valuable towards predicting multiple days into the future than it is towards next-day predictions. Future work could consider how the predictive capabilities change with respect to the distance of the prediction from the input data.

\section{Acknowledgements}
We thank Jillian Wright and Brad Cosma for their contributions to this research, as well as Stephan Sturm, Marcel Blais for their advising. We also thank Branka Hadji-Misheva and J\"org Osterrieder for sponsoring this project and being great mentors and points of assistance. We thank Connor West for his invaluable assistance in debugging. Results in this paper were obtained in part using a high-performance computing system at WPI acquired through NSF MRI grant DMS-1337943.

\bibliographystyle{unsrt}  
\bibliography{references}

\clearpage

\appendix

\section{Definitions}

\subsection{Hyperparameter Search}
Almost all machine learning methods can be tailored to a particular application through the use of hyperparameters. Hyperparameters are the settings and arguments utilized when initializing machine learning models. Available hyperparameters differ significantly depending on the technique being used. The best parameters for a model may vary substantially based on the dataset and prediction task. The correct selection of these hyperparameters can drastically speed up model training and improve performance. Traditionally, a hyperparameter search is performed by selecting a manually inferred set of ``good'' parameters. However, when it is not possible to infer an ideal parameter set, parameter searches are utilized. One very common parameter each technique is grid search, which utilizes an exhaustive grid of acceptable values for each parameter. Grid search generates the set containing all possible combinations of parameters in the provided parameter grid. Models are then fitted on each of these parameter combinations. Each model is ranked based on their performance with respect to a desired metric, and the best parameter set is returned. To improve the reliability of the search, cross-validation is commonly used internally and ranking is done on the mean cross-validation performance. However, grid search can be notably impractical when there is a large number of potential parameters, as the search space grows exponentially. It may also be the case that only a small subset of hyperparameters have a significant impact on model performance. However, this is rarely known for a given model and dataset without performing a search. To remedy this, there have been attempts to create methods that would algorithmically determine the best hyperparameters without having to manually specify a grid to search through~\cite{claesenHyperparameterSearchMachine2015}. 

\subsection{Decision Tree}
Decision Trees are a commonly used technique for explaining the relationships between data and certain outcomes. To do so, they make use of sequences of boolean operations in a tree-like fashion to model a relationship. Specifically, fitting a decision tree model generates a search tree where each node (split) in the tree represents a boolean operation, such that passing the operation induces rightward traversal, while failing the operation induces leftward traversal. When input data is numeric in nature, most (if not all) boolean operations in the tree are less-than comparisons (e.g., ``is the opening price less than \$300?'' Move right if true, left if false). The endpoint of each path of the tree, called leaf nodes, specify a prediction value. In the case of classification problems, each leaf node will be given a value representing some class. Note that multiple leaf nodes may denote the same class, allowing for highly complex decision boundaries. Altogether, this structure makes decision trees very easily explainable models, allowing them to be used both as predictive and descriptive models.

Decision trees are fitted by splitting the provided training dataset based on a boolean operation which maximize the calculated quality of the split. The two most commonly used metrics of split quality for classification are Gini Impurity and Entropy (aka Log-Loss). This process is then repeated recursively for each newly created node and their corresponding subsets of the original dataset. Most frequently, the technique used to implement this is a greedy algorithm. Recursive splitting down a branch of the tree may end when some criteria is met, such as reaching the maximum acceptable depth, splitting the tree again is no longer valuable, the proportion of the dataset represented by a leaf node is below some threshold, or all inputs at a node have the same output value. This makes decision trees highly customizable in their complexity, without sacrificing their explainability. The downside to this is that decision trees are highly prone to overfitting if made too complex, while also potentially being ineffective if made too simple, requiring the implementer to strike a fine balance in complexity to build a performant model~\cite{nandiConditionMonitoringVibration2019}.

\subsection{Random Forest}
Random forest is an ensemble technique which use decision trees. This means that multiple decision trees are fitted and utilized internally to generate a consensus in order to make a prediction. Random forest models use bootstrap aggregation (also known as ``bagging'') to fit a large set of (ideally) uncorrelated decision trees. The intent with this technique is that while the individual decision trees are not highly correlated with each other, the majority of them will still come to the same (correct) conclusion. This mitigates the mistakes of individual decision trees. The downside to this technique is that random forests are highly resource-intensive to train. Further, it is much harder to explain random forest models compared to individual decision trees, as a single random forest model can consist of a massive number of models, each coming to different conclusions~\cite{nandiConditionMonitoringVibration2019}.

\subsection{Logistic Regression}
Logistic regression is commonly used for classification problems where the target variable fits into one of two classes. To do this, logistic regression utilizes log odds, or logit---the probability of success divided by probability of failure---are used for prediction to identify a probability for a given outcome for some input. To make accurate predictions, a weight coefficient $\beta$ for the logistic regression equation is estimated using the maximum likelihood equation. The equation is optimized by testing different $\beta$ values through multiple iterations to see which provides optimal results with the log odds equation. Varying the value of $\beta$ allows us to compute a log-likelihood function, our goal with which is to maximize, theoretically providing with the best value of $\beta$ for prediction. The optimized model can then be used to calculate a probability for a particular class given some input~\cite{nandiConditionMonitoringVibration2019}. 

\subsection{Support Vector Machine}
Support vector machines (SVM), are a type of supervised classification algorithm. SVM can be used for classification, regression, and outlier detection tasks. By default, most SVM models are binary classifiers. To perform this task, SVM attempts to find a decision boundary that optimally separates data points based on their class to maximize the distance between opposing classifications~\cite{alma9936924195604746}. The support vectors are defined as the points close to the decision boundary, as these influence the hyperplane's position and orientation the most. Unlike logistic regression, where output values are scaled to fit in the 0 to 1 range (inclusive), SVM scales the values to -1 to 1 range (inclusive). SVMs also have the option of being fitted using a variety of kernels, enabling polynomial boundaries or multi-class classification. Notable types of kernels include radial basis function, polynomial, sigmoid, and linear. Regularization can also be applied at the kernel level to reduce bias and prevent overfitting. Each type of kernel has its own parameters---such as the degree of the polynomial for the polynomial kernel---thus requiring hyperparameter searches in many cases~\cite{alma9936924195604746}.

\subsection{K-Nearest Neighbors}
K-Nearest Neighbors is a commonly used machine learning technique which utilizes a consensus of similar results to make predictions. At its core, k-nearest neighbors identifies the $k$ closest known samples to the current input and uses them to inform its decision. For classification, the output is chosen as whichever class is most common among the identified neighbors. For regression, the output is the mean value of the selected neighbors. With low values of $k$, small clusters of classes or values can be identified, but overfitting may occur more often. Larger values of $k$ are better at mitigating overfitting and generalizing on a large dataset, but may not be able to account for micro-clusters, enclaves inside the space of another class, or sudden changes in the distribution of values~\cite{kramerKNearestNeighbors2013}.

\subsection{Accuracy and F1 Score}
Accuracy measures how frequently a model predicts correctly as a percentage of the overall provided data. F1 score measures model performance by comparing the model's precision (true positives divided by all positive predictions) and recall (true positives divided by all positive values in the dataset). F1 is slightly better than accuracy, as it also considers the impact of incorrect predictions and the types of errors observed. F1 scores are averaged by class using oen of two strategies: Macro-averaging and weighted averaging. Macro F1 is calculated by summing all class-specific F1 scores and dividing that total by the number of classes. Weighted F1 score takes into account the number of samples of each class, multiplying each class's F1 score by the class's proportion of the dataset.

\subsection{Area Under Receiver Operator Characteristic Curve} 
The Receiver Operator Characteristic curve describes the predictive capabilities of a model as a function of its true positive rate versus its false positive rate. This relationship is generated by varying the probability thresholds used to make predictions. The area under this curve (AUC) describes the ability of a model to separate the desired classes. I.e., a greater area under the curve indicates a better model, where a value of 1 indicates a perfect model.

\newpage

\section{Full Model ANOVA Tables}\label{app:full anova models}

\begin{table}[htbp]
\centering
\caption{Full and reduced ANOVA models for decision tree performance}
\resizebox{\textwidth}{!}{\begin{tabular}{|c||l|rc|rc|rc|rc|}
\hline
 &  & \multicolumn{2}{|c|}{Accuracy} & \multicolumn{2}{|c|}{Macro F1} & \multicolumn{2}{|c|}{Weighted F1} & \multicolumn{2}{|c|}{ROC AUC} \\
Model & Effect & Coefficient & P-value & Coefficient & P-value & Coefficient & P-value & Coefficient & P-value \\
\hline\hline
\multirow[c]{5}{*}{\rotatebox[origin=c]{90}{Full}} & F & $-0.00023$ & $0.785$ & $-0.00018$ & $0.837$ & $-0.00025$ & $0.765$ & $-0.00165$ & $0.075$ \\
 & B & $0.00152$ & $0.067$ & $0.00149$ & $0.086$ & $0.00190$ & $0.021$ & $0.00074$ & $0.422$ \\
 & I & $0.00264$ & $0.001$ & $0.00098$ & $0.259$ & $0.00159$ & $0.053$ & $0.00166$ & $0.073$ \\
 & C & $-0.00141$ & $0.088$ & $-0.00206$ & $0.018$ & $-0.00077$ & $0.348$ & $-0.00237$ & $0.011$ \\ \cdashline{2-10}
 & Intercept & $0.51067$ & - & $0.49941$ & - & $0.50829$ & - & $0.50217$ & - \\ \hline\hline
\multirow[c]{5}{*}{\rotatebox[origin=c]{90}{Reduced}} & F & - & - & - & - & - & - & - & - \\
 & B & - & - & - & - & $0.00190$ & $0.021$ & - & - \\
 & I & $0.00264$ & $0.002$ & - & - & - & - & - & - \\
 & C & - & - & $-0.00206$ & $0.018$ & - & - & $-0.00237$ & $0.011$ \\ \cdashline{2-10}
 & Intercept & $0.51067$ & - & $0.49941$ & - & $0.50829$ & - & $0.50217$ & - \\
\hline
\end{tabular}}
\end{table}

\begin{table}[htbp]
\centering
\caption{Full and reduced ANOVA models for Random Forest performance}
\resizebox{\textwidth}{!}{\begin{tabular}{|c||l|rc|rc|rc|rc|}
\hline
 &  & \multicolumn{2}{|c|}{Accuracy} & \multicolumn{2}{|c|}{Macro F1} & \multicolumn{2}{|c|}{Weighted F1} & \multicolumn{2}{|c|}{ROC AUC} \\
Model & Effect & Coefficient & P-value & Coefficient & P-value & Coefficient & P-value & Coefficient & P-value \\
\hline\hline
\multirow[c]{5}{*}{\rotatebox[origin=c]{90}{Full}} & F & $-0.00097$ & $0.164$ & $-0.00314$ & $0.000$ & $-0.00260$ & $0.000$ & $-0.00204$ & $0.014$ \\
 & B & $0.00457$ & $0.000$ & $0.00096$ & $0.235$ & $0.00228$ & $0.002$ & $0.00229$ & $0.006$ \\
 & I & $0.00051$ & $0.469$ & $-0.00286$ & $0.000$ & $-0.00193$ & $0.010$ & $-0.00224$ & $0.007$ \\
 & C & $0.00225$ & $0.001$ & $-0.00211$ & $0.009$ & $0.00116$ & $0.118$ & $-0.00227$ & $0.007$ \\ \cdashline{2-10}
 & Intercept & $0.53550$ & - & $0.48815$ & - & $0.50863$ & - & $0.50551$ & - \\ \hline\hline
\multirow[c]{5}{*}{\rotatebox[origin=c]{90}{Reduced}} & F & - & - & $-0.00314$ & $0.000$ & $-0.00260$ & $0.000$ & $-0.00204$ & $0.014$ \\
 & B & $0.00457$ & $0.000$ & - & - & $0.00228$ & $0.002$ & $0.00229$ & $0.006$ \\
 & I & - & - & $-0.00286$ & $0.000$ & $-0.00193$ & $0.010$ & $-0.00224$ & $0.007$ \\
 & C & $0.00225$ & $0.001$ & $-0.00211$ & $0.009$ & - & - & $-0.00227$ & $0.007$ \\ \cdashline{2-10}
 & Intercept & $0.53550$ & - & $0.48815$ & - & $0.50863$ & - & $0.50551$ & - \\
\hline
\end{tabular}}
\end{table}

\begin{table}[htbp]
\centering
\caption{Full and reduced ANOVA models for logistic regression performance}
\resizebox{\textwidth}{!}{\begin{tabular}{|c||l|rc|rc|rc|rc|}
\hline
 &  & \multicolumn{2}{|c|}{Accuracy} & \multicolumn{2}{|c|}{Macro F1} & \multicolumn{2}{|c|}{Weighted F1} & \multicolumn{2}{|c|}{ROC AUC} \\
Model & Effect & Coefficient & P-value & Coefficient & P-value & Coefficient & P-value & Coefficient & P-value \\
\hline\hline
\multirow[c]{5}{*}{\rotatebox[origin=c]{90}{Full}} & F & $-0.00661$ & $0.000$ & $-0.00139$ & $0.046$ & $-0.00236$ & $0.000$ & $-0.00947$ & $0.000$ \\
 & B & $-0.00263$ & $0.000$ & $0.00811$ & $0.000$ & $0.00544$ & $0.000$ & $0.00183$ & $0.006$ \\
 & I & $0.01042$ & $0.000$ & $0.01805$ & $0.000$ & $0.01591$ & $0.000$ & $0.02094$ & $0.000$ \\
 & C & $0.00335$ & $0.000$ & $0.00928$ & $0.000$ & $0.00926$ & $0.000$ & $0.00382$ & $0.000$ \\ \cdashline{2-10}
 & Intercept & $0.53654$ & - & $0.51143$ & - & $0.52512$ & - & $0.52818$ & - \\ \hline\hline
\multirow[c]{5}{*}{\rotatebox[origin=c]{90}{Reduced}} & F & $-0.00661$ & $0.000$ & $-0.00139$ & $0.046$ & $-0.00236$ & $0.000$ & $-0.00947$ & $0.000$ \\
 & B & $-0.00263$ & $0.000$ & $0.00811$ & $0.000$ & $0.00544$ & $0.000$ & $0.00183$ & $0.006$ \\
 & I & $0.01042$ & $0.000$ & $0.01805$ & $0.000$ & $0.01591$ & $0.000$ & $0.02094$ & $0.000$ \\
 & C & $0.00335$ & $0.000$ & $0.00928$ & $0.000$ & $0.00926$ & $0.000$ & $0.00382$ & $0.000$ \\ \cdashline{2-10}
 & Intercept & $0.53654$ & - & $0.51143$ & - & $0.52512$ & - & $0.52818$ & - \\
\hline
\end{tabular}}
\end{table}

\begin{table}[htbp]
\centering
\caption{Full and reduced ANOVA models for linear SVM performance}
\resizebox{\textwidth}{!}{\begin{tabular}{|c||l|rc|rc|rc|rc|}
\hline
 &  & \multicolumn{2}{|c|}{Accuracy} & \multicolumn{2}{|c|}{Macro F1} & \multicolumn{2}{|c|}{Weighted F1} & \multicolumn{2}{|c|}{ROC AUC} \\
Model & Effect & Coefficient & P-value & Coefficient & P-value & Coefficient & P-value & Coefficient & P-value \\
\hline\hline
\multirow[c]{5}{*}{\rotatebox[origin=c]{90}{Full}} & F & $-0.00090$ & $0.000$ & $-0.00087$ & $0.188$ & $-0.00079$ & $0.160$ & $-0.00222$ & $0.000$ \\
 & B & $0.00335$ & $0.000$ & $0.01900$ & $0.000$ & $0.01730$ & $0.000$ & $0.00685$ & $0.000$ \\
 & I & $0.00227$ & $0.000$ & $0.00858$ & $0.000$ & $0.00762$ & $0.000$ & $0.01325$ & $0.000$ \\
 & C & $0.00886$ & $0.000$ & $0.00952$ & $0.000$ & $0.01375$ & $0.000$ & $0.00447$ & $0.000$ \\ \cdashline{2-10}
 & Intercept & $0.56966$ & - & $0.40082$ & - & $0.44272$ & - & $0.53837$ & - \\ \hline\hline
\multirow[c]{5}{*}{\rotatebox[origin=c]{90}{Reduced}} & F & $-0.00090$ & $0.000$ & - & - & - & - & $-0.00222$ & $0.000$ \\
 & B & $0.00335$ & $0.000$ & $0.01900$ & $0.000$ & $0.01730$ & $0.000$ & $0.00685$ & $0.000$ \\
 & I & $0.00227$ & $0.000$ & $0.00858$ & $0.000$ & $0.00762$ & $0.000$ & $0.01325$ & $0.000$ \\
 & C & $0.00886$ & $0.000$ & $0.00952$ & $0.000$ & $0.01375$ & $0.000$ & $0.00447$ & $0.000$ \\ \cdashline{2-10}
 & Intercept & $0.56966$ & - & $0.40082$ & - & $0.44272$ & - & $0.53837$ & - \\
\hline
\end{tabular}}
\end{table}

\begin{table}[htbp]
\centering
\caption{Full and reduced ANOVA models for k-nearest neighbors performance}
\resizebox{\textwidth}{!}{\begin{tabular}{|c||l|rc|rc|rc|rc|}
\hline
 &  & \multicolumn{2}{|c|}{Accuracy} & \multicolumn{2}{|c|}{Macro F1} & \multicolumn{2}{|c|}{Weighted F1} & \multicolumn{2}{|c|}{ROC AUC} \\
Model & Effect & Coefficient & P-value & Coefficient & P-value & Coefficient & P-value & Coefficient & P-value \\
\hline\hline
\multirow[c]{5}{*}{\rotatebox[origin=c]{90}{Full}} & F & $-0.01109$ & $0.000$ & $-0.01420$ & $0.000$ & $-0.01315$ & $0.000$ & $-0.01288$ & $0.000$ \\
 & B & $0.00278$ & $0.000$ & $0.00479$ & $0.000$ & $0.00482$ & $0.000$ & $0.00470$ & $0.000$ \\
 & I & $0.00044$ & $0.485$ & $0.00069$ & $0.275$ & $0.00090$ & $0.110$ & $-0.00127$ & $0.050$ \\
 & C & $0.00867$ & $0.000$ & $0.00484$ & $0.000$ & $0.00707$ & $0.000$ & $0.00787$ & $0.000$ \\ \cdashline{2-10}
 & Intercept & $0.52211$ & - & $0.50128$ & - & $0.51397$ & - & $0.51297$ & - \\ \hline\hline
\multirow[c]{5}{*}{\rotatebox[origin=c]{90}{Reduced}} & F & $-0.01109$ & $0.000$ & $-0.01420$ & $0.000$ & $-0.01315$ & $0.000$ & $-0.01288$ & $0.000$ \\
 & B & $0.00278$ & $0.000$ & $0.00479$ & $0.000$ & $0.00482$ & $0.000$ & $0.00470$ & $0.000$ \\
 & I & - & - & - & - & - & - & - & - \\
 & C & $0.00867$ & $0.000$ & $0.00484$ & $0.000$ & $0.00707$ & $0.000$ & $0.00787$ & $0.000$ \\ \cdashline{2-10}
 & Intercept & $0.52211$ & - & $0.50128$ & - & $0.51397$ & - & $0.51297$ & - \\
\hline
\end{tabular}}
\end{table}

\begin{table}[htbp]
\centering
\caption{Full and reduced ANOVA models for random baseline performance}
\resizebox{\textwidth}{!}{\begin{tabular}{|c||l|rc|rc|rc|rc|}
\hline
 &  & \multicolumn{2}{|c|}{Accuracy} & \multicolumn{2}{|c|}{Macro F1} & \multicolumn{2}{|c|}{Weighted F1} & \multicolumn{2}{|c|}{ROC AUC} \\
Model & Effect & Coefficient & P-value & Coefficient & P-value & Coefficient & P-value & Coefficient & P-value \\
\hline\hline
\multirow[c]{5}{*}{\rotatebox[origin=c]{90}{Full}} & F & $0.00000$ & $0.996$ & $0.00000$ & $0.997$ & $0.00000$ & $0.995$ & $-0.00000$ & $0.000$ \\
 & B & $0.00869$ & $0.000$ & $0.00842$ & $0.000$ & $0.00873$ & $0.000$ & $0.00000$ & $0.000$ \\
 & I & $-0.00061$ & $0.451$ & $-0.00090$ & $0.267$ & $-0.00065$ & $0.413$ & $0.00000$ & $0.914$ \\
 & C & $0.00897$ & $0.000$ & $0.00814$ & $0.000$ & $0.00936$ & $0.000$ & $-0.00000$ & $0.596$ \\ \cdashline{2-10}
 & Intercept & $0.51636$ & - & $0.51229$ & - & $0.51819$ & - & $0.50000$ & - \\ \hline\hline
\multirow[c]{5}{*}{\rotatebox[origin=c]{90}{Reduced}} & F & - & - & - & - & - & - & - & - \\
 & B & $0.00869$ & $0.000$ & $0.00842$ & $0.000$ & $0.00873$ & $0.000$ & - & - \\
 & I & - & - & - & - & - & - & - & - \\
 & C & $0.00897$ & $0.000$ & $0.00814$ & $0.000$ & $0.00936$ & $0.000$ & - & - \\ \cdashline{2-10}
 & Intercept & $0.51636$ & - & $0.51229$ & - & $0.51819$ & - & $0.50000$ & - \\
\hline
\end{tabular}}
\end{table}

\begin{table}[htbp]
\centering
\caption{Full and reduced ANOVA models for constant baseline performance}
\resizebox{\textwidth}{!}{\begin{tabular}{|c||l|rc|rc|rc|rc|}
\hline
 &  & \multicolumn{2}{|c|}{Accuracy} & \multicolumn{2}{|c|}{Macro F1} & \multicolumn{2}{|c|}{Weighted F1} & \multicolumn{2}{|c|}{ROC AUC} \\
Model & Effect & Coefficient & P-value & Coefficient & P-value & Coefficient & P-value & Coefficient & P-value \\
\hline\hline
\multirow[c]{5}{*}{\rotatebox[origin=c]{90}{Full}} & F & $0.00001$ & $0.884$ & $0.00001$ & $0.882$ & $0.00002$ & $0.885$ & $0.00000$ & $0.000$ \\
 & B & $0.00180$ & $0.000$ & $0.00074$ & $0.000$ & $0.00212$ & $0.000$ & $0.00000$ & $0.118$ \\
 & I & $0.00041$ & $0.000$ & $0.00016$ & $0.000$ & $0.00050$ & $0.000$ & $0.00000$ & $0.000$ \\
 & C & $0.00736$ & $0.000$ & $0.00300$ & $0.000$ & $0.00872$ & $0.000$ & $-0.00000$ & $0.051$ \\ \cdashline{2-10}
 & Intercept & $0.56610$ & - & $0.36146$ & - & $0.40930$ & - & $0.50000$ & - \\ \hline\hline
\multirow[c]{5}{*}{\rotatebox[origin=c]{90}{Reduced}} & F & - & - & - & - & - & - & $0.00000$ & $0.026$ \\
 & B & $0.00180$ & $0.000$ & $0.00074$ & $0.000$ & $0.00212$ & $0.000$ & - & - \\
 & I & $0.00041$ & $0.000$ & $0.00016$ & $0.000$ & $0.00050$ & $0.000$ & - & - \\
 & C & $0.00736$ & $0.000$ & $0.00300$ & $0.000$ & $0.00872$ & $0.000$ & - & - \\ \cdashline{2-10}
 & Intercept & $0.56610$ & - & $0.36146$ & - & $0.40930$ & - & $0.50000$ & - \\
\hline
\end{tabular}}
\end{table}

\begin{table}[htbp]
\centering
\caption{Full and reduced ANOVA models for previous baseline performance}
\begin{tabular}{|c||l|rc|rc|rc|}
\hline
 &  & \multicolumn{2}{|c|}{Accuracy} & \multicolumn{2}{|c|}{Macro F1} & \multicolumn{2}{|c|}{Weighted F1} \\
Model & Effect & Coefficient & P-value & Coefficient & P-value & Coefficient & P-value \\
\hline\hline
\multirow[c]{5}{*}{\rotatebox[origin=c]{90}{Full}} & F & $0.00006$ & $0.869$ & $0.00006$ & $0.860$ & $0.00006$ & $0.861$ \\
 & B & $0.01519$ & $0.000$ & $0.01500$ & $0.000$ & $0.01516$ & $0.000$ \\
 & I & $0.00929$ & $0.000$ & $0.00931$ & $0.000$ & $0.00927$ & $0.000$ \\
 & C & $0.00414$ & $0.000$ & $0.00229$ & $0.000$ & $0.00410$ & $0.000$ \\ \cdashline{2-8}
 & Intercept & $0.53557$ & - & $0.52733$ & - & $0.53549$ & - \\ \hline\hline
\multirow[c]{5}{*}{\rotatebox[origin=c]{90}{Reduced}} & F & - & - & - & - & - & - \\
 & B & $0.01519$ & $0.000$ & $0.01500$ & $0.000$ & $0.01516$ & $0.000$ \\
 & I & $0.00929$ & $0.000$ & $0.00931$ & $0.000$ & $0.00927$ & $0.000$ \\
 & C & $0.00414$ & $0.000$ & $0.00229$ & $0.000$ & $0.00410$ & $0.000$ \\ \cdashline{2-8}
 & Intercept & $0.53557$ & - & $0.52733$ & - & $0.53549$ & - \\
\hline
\end{tabular}
\end{table}

\begin{table}[htbp]
\centering
\caption{Full and reduced ANOVA models for consensus baseline performance}
\begin{tabular}{|c||l|rc|rc|rc|}
\hline
 &  & \multicolumn{2}{|c|}{Accuracy} & \multicolumn{2}{|c|}{Macro F1} & \multicolumn{2}{|c|}{Weighted F1} \\
Model & Effect & Coefficient & P-value & Coefficient & P-value & Coefficient & P-value \\
\hline\hline
\multirow[c]{5}{*}{\rotatebox[origin=c]{90}{Full}} & F & $0.00006$ & $0.761$ & $0.00006$ & $0.683$ & $0.00005$ & $0.808$ \\
 & B & $-0.00157$ & $0.000$ & $-0.00320$ & $0.000$ & $-0.00112$ & $0.000$ \\
 & I & $-0.00145$ & $0.000$ & $-0.00110$ & $0.000$ & $-0.00185$ & $0.000$ \\
 & C & $0.00855$ & $0.000$ & $0.00537$ & $0.000$ & $0.00889$ & $0.000$ \\ \cdashline{2-8}
 & Intercept & $0.49597$ & - & $0.47412$ & - & $0.50353$ & - \\ \hline\hline
\multirow[c]{5}{*}{\rotatebox[origin=c]{90}{Reduced}} & F & - & - & - & - & - & - \\
 & B & $-0.00157$ & $0.000$ & $-0.00320$ & $0.000$ & $-0.00112$ & $0.000$ \\
 & I & $-0.00145$ & $0.000$ & $-0.00110$ & $0.000$ & $-0.00185$ & $0.000$ \\
 & C & $0.00855$ & $0.000$ & $0.00537$ & $0.000$ & $0.00889$ & $0.000$ \\ \cdashline{2-8}
 & Intercept & $0.49597$ & - & $0.47412$ & - & $0.50353$ & - \\
\hline
\end{tabular}
\end{table}

\end{document}